\documentclass[prl,twocolumn]{revtex4}
\usepackage{graphicx}
\usepackage{dcolumn}
\usepackage{bm}
\usepackage{amssymb,amsmath}

\begin{document}
\title {Highly enhanced thermopower in
two-dimensional electron systems at milliKelvin temperatures}
\author{Srijit Goswami}
\email{sg483@cam.ac.uk}\affiliation{Cavendish Laboratory, University of
Cambridge, J.J. Thomson Avenue, Cambridge CB3 0HE, United Kingdom.}
\author{Christoph Siegert}
\affiliation{Cavendish Laboratory, University of Cambridge, J.J. Thomson
Avenue, Cambridge CB3 0HE, United Kingdom.}
\author{Matthias Baenninger}
\affiliation{Cavendish Laboratory, University of Cambridge, J.J. Thomson
Avenue, Cambridge CB3 0HE, United Kingdom.}
\author{Arindam Ghosh}
\email{arindam@physics.iisc.ernet.in} \affiliation{Department of Physics,
Indian Institute of Science, Bangalore 560 012, India.}
\author{Michael Pepper}
\thanks{Present address: Department of Electronic and Electrical Engineering, University College, London.}
\affiliation{Cavendish Laboratory, University of Cambridge, J.J. Thomson
Avenue, Cambridge CB3 0HE, United Kingdom.}
\author{Ian Farrer}
\affiliation{Cavendish Laboratory, University of Cambridge, J.J. Thomson
Avenue, Cambridge CB3 0HE, United Kingdom.}
\author{David A. Ritchie}
\affiliation{Cavendish Laboratory, University of Cambridge, J.J. Thomson
Avenue, Cambridge CB3 0HE, United Kingdom.}


\begin{abstract}
We report experimental observation of an unexpectedly large thermopower in
mesoscopic two-dimensional (2D) electron systems on GaAs/AlGaAs
heterostructures at sub-Kelvin temperatures and zero magnetic field. Unlike conventional non-magnetic high-mobility
2D systems, the thermopower in our devices increases with decreasing
temperature below 0.3~K, reaching values in excess of
100~$\mu$V/K, thus exceeding the free electron estimate by more than two orders of magnitude. With support from a parallel independent study of the local
density of states, we suggest such a phenomenon to be linked to intrinsic
localized states and many-body spin correlations in the system.
\end{abstract}


\maketitle

The diffusion thermopower (TP), $S_d$, in a solid depends on the variation of
scattering time or the density-of-states (DOS) in the vicinity of the Fermi
energy ($E_F$), and expressed by the Mott
relation~\cite{mottjones, cutlermott}:

\begin{equation}
\label{eq1} S_{d} = \lim_{\Delta T\rightarrow 0} \frac{V_{th}}{\Delta T} =
-\frac{\pi^2k_B^2T}{3|e|} \frac{d\ln \sigma(E)}{dE}\Big |_{E=E_F}
\end{equation}

\noindent where $V_{th}$ is the thermovoltage at temperature $T$, and
$\sigma(E)$ is the energy-dependent conductivity. Consequently, TP is directly
sensitive to proximity to band edge or gaps in DOS in strongly localized
systems. This has been exploited in investigating the localization transition
in a disordered 2DES~\cite{mitmosfet:fletcher}, where the 2DES goes from a purely metallic state to a highly localized one. Similar TP measurements have been used to explore quantum
insulating ground states of a bulk 2D electron gas in high magnetic fields~\cite{holegasQH:possan}. On
the other hand, being a measure of entropy per carrier, the third law of
thermodynamics requires TP $\rightarrow 0$ at $T \rightarrow 0$ in delocalized
or metallic systems, although the absolute magnitude of TP depends critically
on the energy-dependence of scattering mechanism of the conduction electrons.
This has direct implications for a large classs of highly correlated 3D systems,~\cite{kondoce1:kowal, kondoce2:garde,
kondoce3:pinto,kondoyb1:akrap, kondoybce:zlatic, kondoHFtheory:Schweitzer},
such as the dilute magnetic alloys, heavy Fermions or Kondo lattice compounds. 
Here, the conduction electrons undergo spin-flip scattering at the localized $d-$
or $f-$ sites, which may lead to very large TP, exceeding several hundred
$\mu$V/K~\cite{largekondofe, largekondo:harut}. In this study, we focus on nonmagnetic
delocalized high-mobility 2D electron systems (2DES) at GaAs/AlGaAs interface,
which are not expected to contain localized spins, and indeed, earlier
measurements of zero magnetic field diffusion TP in macroscopic 2DES yielded TP
$\lesssim 0.4~\mu$V/K even at $T$ as large as $\thickapprox
3$~K~\cite{maximov}.

Recently though, several experiments on nonequilibrium transport in quasi-one
dimensional quantum point contacts (QPC)~\cite{pointseven:KJT,pointseven:cronen} and unconfined mesoscopic 2DES \cite{spinpol:ghosh,kondo:ghosh,lattice:siegert} suggest the
possibility of localized spins in these III-V semiconductor-based
low-dimensional systems. While the microscopic origin of the localized spins in
such materials remains unclear, measurements in QPCs have indeed revealed
additional contribution to the diffusion thermopower near the so-called '0.7 state' that is not
entirely understood~\cite{pointseven:appleyard}. A more controlled study has
been carried out with electrostatically-confined single localized spins in
odd-electron quantum dots, where the spin-flip scattering was found to result in large TP ($\simeq 60~ \mu$V/K) at low $T$~\cite{Kondo:Scheibner}. The enhanced
TP was attributed to spin-entropy transfer, similar to that in layered cobalt
oxides~\cite{Cobalt}. Although GaAs/AlGaAs-based high mobility 2DESs form the
host in many of these studies, no systematic TP experiments have been reported
in unconfined 2DESs at mesoscopic length scales, which could provide crucial insight into the nature and role of intrinsic spins. Here we
report experimental observation of a giant diffusion TP in delocalized
(conductivity $\gg e^2/h$) 2DES in high-mobility GaAs/AlGaAs heterostructures
at T $<$ 0.3~K. We find that at most gate voltages
$|S_{d}|$ {\it increases} with decreasing $T$ down to the base lattice
temperature ($T_{latt}$) of $\simeq 70$~mK, indicating that it cannot be
explained with free noninteracting electrons in 2D. We complement thermal measurements with conventional nonequilibrium
transport, and show that these results point strongly towards the existence of
localized spins in 2D mesoscopic systems. Our findings may have a direct impact
on the understanding of many experimentally reported, but not fully understood,
phenomena in low-dimensional quantum systems, such as the `0.7-state'
\cite{pointseven:KJT} and breakdown of Wiedemann-Franz law in quantum point
contacts (QPC)~\cite{wf:chiatti}, the zero-bias
anomaly~\cite{lattice:siegert, spinpol:ghosh, kondo:ghosh}, or the anomalous Hall
effect in 2DESs at ultra-low temperatures~\cite{hall:siegert}.

\begin{figure}[h]
\includegraphics[width=1\linewidth]{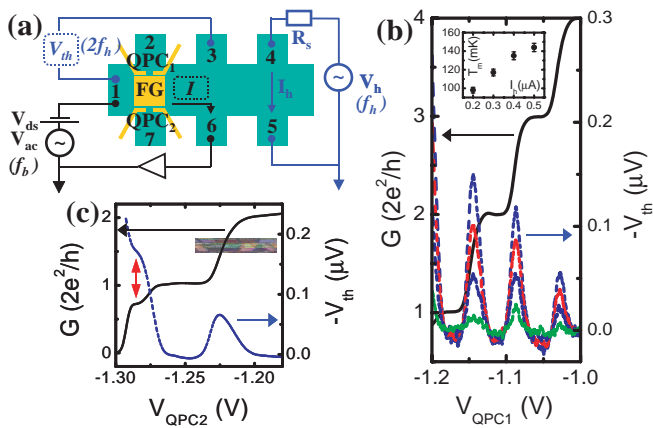}
\caption{(Color online) (a) Schematic of the experimental setup used to perform thermovoltage (blue lines) and conductance (black lines) measurements.  (b)
Characteristic peaks in $V_{th}$ as a function of the QPC gate voltage
($V_{QPC1}$) for heating currents, $I_{h}$, ranging from 0.2 $\mu$A (bottom
trace) 0.5 $\mu $A (topmost trace). (Inset) Electron temperature
at the center of the 2D mesoscopic region obtained from the QPC TP analysis (see text). (c) A similar trace for $QPC_2$,
where a distinct 0.7 structure was observed. Arrows mark the position of the
0.7 structure in conductance and the corresponding deviation of the
thermovoltage from the semiclassical expectation. (Inset) Scanning
electron micrograph of the device with a central $5\mu m\times 5\mu m$ gate
used to form the mesoscopic system.}
\end{figure}
Silicon delta-doped heterostructures with thick ($\approx 80$~nm) spacer layer
of undoped AlGaAs (to minimize remote impurity scattering) were employed in our
experiments. Similar devices were earlier used for nonequilibrium transport and
Hall measurements~\cite{hall:siegert, spinpol:ghosh, kondo:ghosh}. The mobility of the
electrons in these wafers were found to be in the range of $1 -
3\times10^6$~cm$^2$/V-s at the as-grown electron sheet density, $n_s \simeq 1\times
10^{11}$cm$^2$ resulting in a long elastic mean free path ($\gtrsim 10\mu$m). A
schematic of the device-structure for TP measurements with the gate
assembly is shown in Fig.~1a (see inset of Fig.~1c for the SEM micrograph of
the device region). Central to the design is the $5\mu$m$\times5\mu$m full gate
(FG) which forms the mesoscopic device under study. A voltage ($V_{FG}$) on this
gate tunes $n_s$ in the 2DES directly below, thus allowing for a detailed study
of the thermovoltage as a function of $E_F$ of the mesoscopic system. The QPCs
on either side of FG serve three purposes: (1) Lateral isolation of the
device region from the remaining ungated 2DES, (2) validation of the
measurement technique with known thermoelectric behavior of QPCs, and (3)
electron temperature calibration of the mesoscopic region with respect to
$T_{latt}$  using the procedure described in Ref~\cite{Thermometer:Appleyard}.
An oscillatory heating current ($I_h^{4,5}$) with frequency $f_h = 7.3$~Hz was used
between remote leads 4 and 5. The thermovoltage detection was performed with a
lock-in amplifier at $2f_h$ to ensure a purely thermal origin of the signal.

Fig.~1b shows $-V_{th}^{2,7}$ across $QPC_{1}$ as a function of the split-gate
voltage $V_{QPC1}$ for various $I_h$ ($0.2\mu$A$\rightarrow 0.5\mu$A) with
$T_{latt}$ fixed at the base. The peaks in $|V_{th}|$ between two
consecutive conductance plateaus could be scaled (not shown) for $I_h \leq
0.5\mu$A yielding the absolute electron temperature $T_m$ at the center of the
mesoscopic region as a function of $I_h$ (see inset of Fig.~1b). The quantitative agreement
between measured TP of the QPC and the energy-derivative of its conductance (through Eq. 1)
supports the independent, non-interacting electron description of the QPC at higher subbands.
We also notice that, (1) the TP in our devices has a purely diffusive (and
ballistic for QPC) origin, and any contribution from phonon drag is negligible,
as expected for GaAs-based 2DESs below $\sim 300$~mK \cite{oscill:fletcher}.
(2) Scaling of thermopower for $I_h \leq 0.5\mu$A suggests thermal broadening
to be negligible at these currents. (3) Fig.~1c shows the TP in $QPC_2$
(keeping $V_{QPC1}$ fixed) below the first subband where a clear deviation from
the Mott relation is observed near the '0.7 structure'. This deviation
has been studied by Appleyard \emph{et al.} \cite{pointseven:appleyard}, but
its origin is yet to be completely understood. However it serves as proof that
the observed TP is indeed capable of detecting signatures of many-body
spin-correlated states in low-dimensional systems. For subsequent measurements
of thermopower across the mesoscopic region, both QPCs were pinched off, and
$V^{1,3}_{th}$ was measured after adequate amplification. The
temperature difference ($\Delta T$) across the device was estimated as $\Delta
T \approx (L/\xi)[T_m(I_h) -T_{latt}]$, where $L
(= 5\mu$m) and $\xi (\approx 100\mu$m) are the device length and thermal
relaxation length in high-mobility GaAs/AlGaAs systems \cite{thermomagnetic:pogosov,Thermometer:Appleyard},
respectively. For electrical transport, we have used a standard ac/dc technique
to measure both linear response conductivity ($G^{1,6}(V_{ds} = 0)$) and
non-equilibrium differential conductivity ($dI(V_{ds})/dV$), where $V_{ds}$ is
the drain-source bias. All measurements were carried out at zero magnetic
field. The electrical characteristics were recorded only at $I_h = 0$, the
thermovoltages were measured at $I = 0$.

\begin{figure}[h]
\begin{center}
\includegraphics[width=1\linewidth]{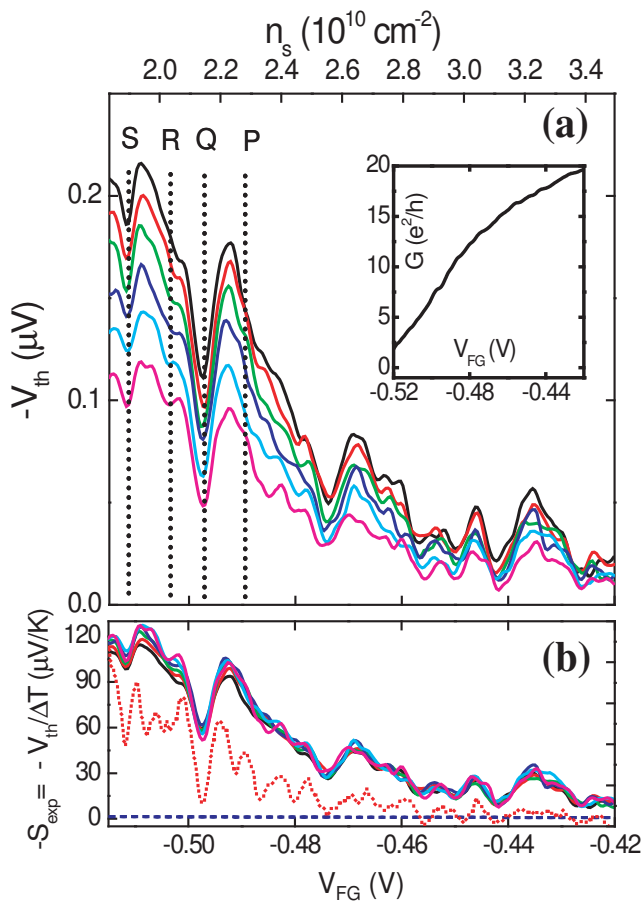}
\caption{(Color online) (a) Variation of $-V_{th}$ as a function of $V_{FG}$ for $I_h$ ranging
from $0.15\mu$A (bottom trace) to $0.25\mu$A (topmost trace). The top axis
shows the corresponding electron sheet density ($n_s$) of the mesoscopic system. (Inset)
Equilibrium conductance ($G$) in this $V_{FG}$ range. (b) Curves in (a) scaled by $\Delta T$ to obtain thermopower ($S_{exp}$). (Dotted) Calculated $d\ln G/dV_{FG}$ using $G$ from Figure 2, confirming the diffusive origin of TP. (Dashed) Expected TP in the free electron picture.}
\end{center}
\end{figure}

Fig.~2a shows the measured thermovoltage for our mesoscopic device at various
$I_h$ as a function of $V_{FG}$ at $T_{latt} \simeq 70$~mK. We concentrate on
the delocalized regime of the 2DES, over $G \sim (2 - 20)\times e^2/h$, which
corresponds to $V_{FG} \gtrsim -0.52$~V, or equivalently, $n_s \gtrsim
1.8\times10^{10}$~cm$^{-2}$ (see inset of Fig.~2a). A reproducible fluctuating
behavior in $V_{th}$ rides on an overall background that increases with
decreasing $n_s$. Note that the decreasing magnitude of $V_{th}$ with
increasing $V_{FG}$ ensures that the contribution to $V_{th}$ from the ungated
part of the 2DES between the thermovoltage leads (1 and 3) is negligible,
and the measured $V_{th}$ arises predominantly from the TP of the mesoscopic
region only. The $V_{th}$ vs. $V_{FG}$ traces could be collapsed onto a single
trace by normalizing each $V_{th}$ with the corresponding $\Delta T$ obtained
from QPC calibration. Strikingly, Fig.~2b indicates the TP of our mesoscopic
device to become $\gtrsim 100 \mu$V/K at low $n_s$, unexpectedly large for a
delocalized system at subKelvin temperatures and zero magnetic field. While the
qualitative agreement with the energy derivative of linear conductivity (red
dotted line) provides further support to the diffusive origin of TP, the
scaling also confirms that increasing $I_h$ does not lead to thermal broadening
or substantial lattice heating for phonon drag to become important.

Increase in TP with decreasing $n_s$ can be envisaged for free degenerate
electrons, as well as for systems close to a localization
transition~\cite{mitmosfet:fletcher, gethermo:burns, pd:burnschaikin}. The latter is plausible at $G
\ll e^2/h$, when the 2DES becomes inhomogeneous deep into the band tail, and transport
is dominated by classical percolation~\cite{percol:sds, droplet:tripathi, pd:burnschaikin}. In our case however,
$G \gg e^2/h$, and direct Hall measurements also indicate the charge
distribution to be uniform~\cite{hall:siegert}. Moreover, the TP never reverses
its sign, ruling out sequential or co-tunneling effects in unintentional
quantum dots within mesoscopic region~\cite{droplet:tripathi, blockade:tripathi}. In the free electron scenario with scattering
from dopant potential, Eq.~\ref{eq1} estimates the TP to be $\approx
-\pi^2k_B^2T(p+1)/3|e|E_F$ ($p \approx 1.5$) \cite{fletcher:highbthermo}, which is nearly two orders of
magnitude lower than the experimentally observed magnitude (dashed line in
Fig.~2b).

\begin{figure}[h]
\begin{center}
\includegraphics[width=1\linewidth]{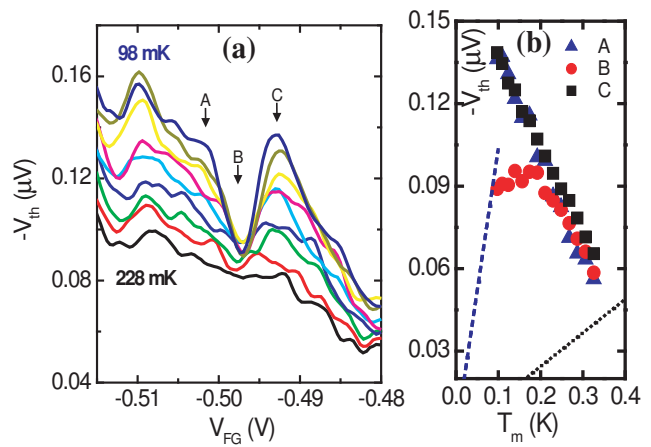}
\caption{(Color online) (a) $-V_{th}$ versus $V_{FG}$ for $98\textrm{ mK} \leq T_{m}\leq
228 \textrm{ mK}$. (b) Distinct $T$-dependences for positions marked A, B and C in
(a). Calculated $T$ dependence expected for a free electron system (dotted) and Kondo lattice system for $T < T_{K}$(dashed).}
\end{center}
\end{figure}

The temperature dependence of $V_{th}$ provides further evidence of nontrivial
origin of the enhanced TP in our devices. This is shown in Fig.~3a for a
selected range of $V_{FG}$ centered around a local minimum at $V_{FG} \approx
-0.497$~V. Increasing $T_m$ from 98~mK to 228~mK (at constant $I_{h}=0.2$ $\mu$A), washes out strong fluctuations
in $V_{th}$, and results in a decrease in the its overall magnitude for $T_m
\gtrsim 150$~mK. Two distinct behaviors were observed (see Fig.~3b): At
$V_{FG}$, labeled A and C, away from the local minimum, $|V_{th}|$
increases monotonically with decreasing $T_m$, but at the minimum, B,
$|V_{th}|$ saturates at an intermediate temperature, and even decreases when
$T_m$ is reduced further. Clearly, this is not expected in the free electron scenario where $|V_{th}| \propto T_m$
(the black dotted line in Fig.~3b), neither in the proximity of localization
where TP is expected to increase with decreasing $T_m$ at all
$V_{FG}$~\cite{gethermo:burns, mitmosfet:fletcher, pd:burnschaikin}.

\begin{figure}[h]
\begin{center}
\includegraphics[width=1\linewidth]{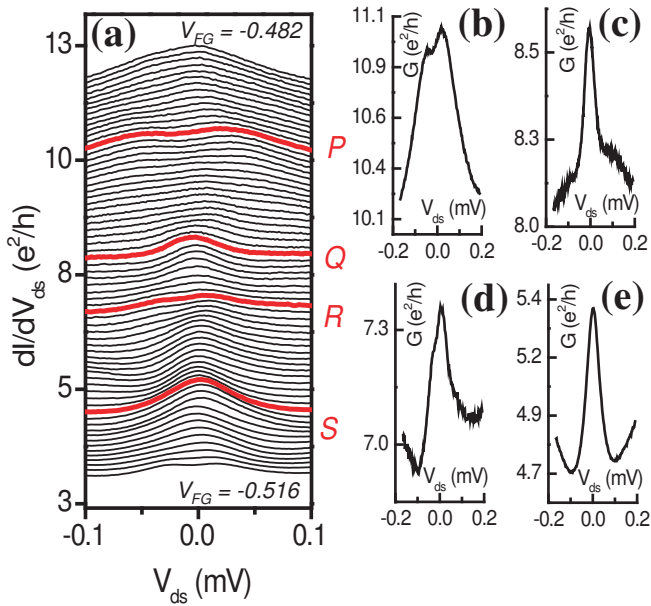}
\caption{(Color online) Non-equilibrium transport (a) Continuous variation of the nature of
the ZBA with $V_{FG}$. Bold plots correspond to $V_{FG}$ values marked P-S in
Figure 2a. P,R(Q,S) show the strongest double (single) peaked ZBAs.  (b)-(d)
Individual ZBA corresponding to P-S in (a).}
\end{center}
\end{figure}

The feasibility of spontaneous spin effects in high-mobility non-magnetic 2DESs
have been discussed widely~\cite{spinpol:zozo, spinpol:ghosh, lattice:siegert}, where both
the background disorder and many-body exchange interaction play crucial roles.
While the signature of exchange-driven Stoner ferromagnetism on the TP of a
2DES at low $T$ is not clear, a disorder-induced two-component fluctuation in
the conduction band may lead to localized spins embedded inside the delocalized
Fermi sea~\cite{lattice:siegert}, and may form a scenario similar to
Kondo-lattice compounds or dilute magnetic alloys. The gate voltage $V_{FG}$
tunes the Fermi wave-vector, and consequently the relative scales of Kondo
coupling and RKKY magnetic exchange, which have characteristic signatures in the
DOS in the vicinity of $E_F$. In nonequilibrium electrical transport, such low
energy structures in the DOS are manifested as a zero-bias anomaly (ZBA) in
$dI/V_{ds}$ around $V_{ds} = 0$. Fig.~4 shows two different forms of ZBA in our
device as $V_{FG}$ in changed, alternating between a single-peak at $V_{ds} =
0$ representing the Kondo resonance at individual localized spins,  and a
double-peaked ZBA that has a shallow minimum close to the $E_F$ indicating
finite inter-spin interaction and local magnetic
ordering~\cite{lattice:siegert}. We note that both temperature and
$V_{FG}$-dependence of $V_{th}$ are directly connected to the structure of
respective ZBAs. The strong minima observed in $|V_{th}|$ at -0.497 V and
-0.510 V (indicated by S and Q in Figs.~2 and 4) correspond to strong single
peaked ZBAs, {\it i.e.} when inter-spin exchange is small and Kondo-coupling
dominates at low $T$. Analogous to heavy Fermions or dilute magnetic alloys,
the $T$-dependence of $|V_{th}|$ in this state is nonmonotonic~\cite{anderson:cox,kondoHFtheory:Schweitzer}(see trace B in
Fig.~3), and shows a downturn just below $T_K
(\approx 0.2$~K $-0.3$~K. ($T_K$ was determined independently from both equilibrium
and nonequilibrium transport. For details see~\cite{kondo:ghosh}) When
spin-spin exchange is strong, illustrated by traces at P and R with split ZBA,
$|V_{th}|$ increases monotonically down to base temperature, implying that the
giant thermopower in our devices at low temperatures could probably be a result
of scattering of the conduction electrons by
unscreened magnetic moments within the 2DES.

Many quantitative approaches towards thermal transport coefficients in Kondo
lattice systems exist in the periodic Anderson model~\cite{anderson:cox},
although primarily in the context of heavy atom intermetallic
compounds~\cite{kondoce1:kowal, kondoce2:garde, kondoyb1:akrap, kondoce3:pinto,
largekondofe, largekondo:harut}. Nevertheless, in the absence of magnetic
interactions, a universal behavior of TP at $T \ll T_K$ in these systems can
be expressed as $S_d = -\alpha(k_B/|e|)(T/T_{K})$ \cite{anderson:cox}, where $\alpha \sim {\cal
O}[1]$. An estimate with $\alpha = 2$, and typical experimental $T_K = 0.25$~K
is shown as the dashed line in Fig.~3b. Despite uncertainties in the numerics,
the asymptotic behavior of trace B appears encouraging, although measurements
need to be extended to lower electron temperatures for more quantitative
conclusions. Two important points to note: (1) Unlike many Kondo lattice
systems, or even semiconductor quantum dots~\cite{Kondo:Scheibner}, we do not
find any change in sign of $V_{th}$ within the experimental temperature range,
remaining negative throughout. This also indicates that the average energy of
the quasiparticles is negative as would be expected for a quasi-ballistic electronic system in
the Kondo regime~\cite{Kondo:Scheibner}. (2) We find that the modulations in
$V_{th}$ can be traced to $V_{FG}$ as large as $-0.42$~V, where the large
zero-bias conductance ($G(V_{ds} = 0) \approx 20\times e^2/h$) makes the ZBA
essentially undetectable. This establishes the greater sensitivity of the TP
over electrical transport to detect anomalies in the local DOS near $E_F$.

In conclusion, we have measured unexpectedly large values of diffusion
thermopower (in excess of $100 \mu$V/K) in delocalized 2D mesoscopic electron
systems. Below 300~mK, the thermopower was found to increase with decreasing
temperature indicating the failure of non-interacting electron model in
this regime. We suggest that the observed enhancement in
thermopower may be related to the formation of localized spins in the system,
and draw analogies between nonmagnetic high-mobility electron devices and Kondo
lattice compounds.

This work was supported by EPSRC (U.K.) and UK-India Education and Research (UKIERI)
Initiative. SG would like to thank the Gates Cambridge Trust for financial support.


\end{document}